\documentclass[preprint,prb,superscriptaddress,byrevtex,showpacs,
showkeys]{revtex4}%
\usepackage{amsmath}
\usepackage{amsfonts}%
\usepackage{amssymb}%
\usepackage{graphicx}
\providecommand{\U}[1]{\protect\rule{.1in}{.1in}}
\newtheorem{theorem}{Theorem}
\newtheorem{acknowledgement}[theorem]{Acknowledgement}

\begin{document}
\preprint{ }
\title[Short title for running header]{Doping effects in the coupled, two-leg spin ladder BiCu$_{2}$PO$_{6}$}
\author{B. Koteswararao}
\affiliation{Department of Physics, Indian Institute of Technology Bombay, Mumbai 400076, India}
\author{A. V. Mahajan}
\affiliation{Department of Physics, Indian Institute of Technology Bombay, Mumbai 400076, India}
\author{L.K.Alexander}
\affiliation{Laboratoire de Physique des Solids, Universit\'{e} Paris-Sud, 91405 Orsay, France}
\author{J. Bobroff}
\affiliation{Laboratoire de Physique des Solids, Universit\'{e} Paris-Sud, 91405 Orsay, France}

\begin{abstract}
We report preparation, x-ray diffraction, magnetic susceptibility $\chi(T)$
and heat capacity $C_{p}(T)$ measurements on the undoped samples as also
samples with Zn-doped $(S=0)$ at Cu site, Ni-doped $(S=1)$ at Cu site, and
Ca-doped (holes) at Bi site in the coupled two-leg spin ladder system
BiCu$_{2}$PO$_{6}$. While, Zn shows complete solid solubility, Ni could be
doped to about 20\% and Ca to about 15\%. Magnetisation and heat capacity data
in the undoped compound point towards the existence of frustration effects. In
all the samples, the $\chi(T)$ at low temperature increases with doping
content. The Zn-induced susceptibility is smaller than that due to effective
$S=1/2$ moments possibly due to frustrating next-nearest-neighbour
interactions along the leg. \ For Zn content $x$
$>$
$0.01$, $\chi(T)$ deviates from the Curie-law at low temperatures. The
magnetic\ specific heat data $C_{m}(T)$ for the Zn-doped samples show weak
anomalies at low temperature in agreement with $\chi(T)$ behavior. The
anomalies are suggestive of spin freezing at low-$T$. In contrast, prominent
effects are observed\ in $\chi(T)$ and $C_{m}(T)$ on Ni-doped samples. The
zero-field-cooled (ZFC) and field-cooled (FC) $\chi(T)$ data are different
from each other at low temperature unlike that for Zn-doped samples, clearly
indicating a transition to a spin-glass like phase. No anomalies were found in
Ca- or Pb-doped samples.

\end{abstract}
\volumeyear{year}
\volumenumber{number}
\issuenumber{number}
\eid{identifier}
\date[Date text]{date}
\received[Received text]{date}

\revised[Revised text]{date}

\accepted[Accepted text]{date}

\published[Published text]{date}

\startpage{1}
\endpage{2}
\maketitle

\section{\textbf{INTRODUCTION}}

Currently, there is great interest in the magnetic properties of pure and
doped low-dimensional \ Heisenberg antiferromagnetic (HAF) and frustrated spin
systems. Exotic ground states such as valence bond solid (VBS) state in
Haldane chains, spin-dimers, and spin-Peierls systems\cite{A. N. Vasilev} or
the resonating valence bond (RVB) state in chain,
ladder\cite{E.dagottoScience271} or planar systems\cite{pwanderson}$^{,}%
$\cite{sachdevsci2000} have been investigated. A lot of experimental and
theoretical work has been done on the effect of non-magnetic Zn$^{2+}$ $(S=0)
$ and magnetic Ni$^{2+}$\ ($S=1$) impurities on Cu-based systems. When
Cu$^{2+}$ ions in hole-doped, metallic CuO$_{2}$ planes in YBa$_{2}$Cu$_{3}%
$O$_{6+\delta}$\ are substituted with Zn$^{2+}$, nuclear magnetic resonance
(NMR)\ experiments\cite{avmPRL} revealed that local magnetic moments are
induced on neighboring\ Cu sites. Further NMR studies\cite{J. bobroff PRL83}
investigated the effect of impurities on the magnetic correlations and showed
that each vacancy causes a strong antiferromagnetic polarization around its
immediate vicinity.\cite{MHJulien} This phenomenon was\ also observed in
Ni-doped square lattices.\cite{P. Mendels}\ Similar behavior was also found in
Haldane chains\cite{FTedoldi} and ladders. \cite{S.Ohsugi}$^{,} $%
\cite{gbmartin}

In the case of $S=%
\frac12
,$ even-leg spin ladders (which have a spin gap), a transition from the
spin-liquid ground state to an ordered state across a quantum critical point
(QCP) can occur as a function of some tuning parameter such as the interladder
coupling.\cite{sachdevsci2000} Additionally, non-magnetic impurities have been
found to induce disordered magnetism or magnetic long-range-order (LRO) in a
prototypical two-leg ladder such as SrCu$_{2}$O$_{3}.$\cite{MI Larkin}$^{,}%
$\cite{azuma1997} In SrCu$_{2}$O$_{3}$, the exchange interaction along the leg
($J_{1}/k_{B}$) was found to be $1900$ K. The ratio of the rung exchange to
leg exchange $J_{2}/J_{1}$ and the\ spin-gap ($\Delta/k_{B}$) were found to be
$0.5$ and $450$ K, respectively, with only a weak inter-ladder
interaction.\cite{DC1996} In this spin gapped system, even a\ small amount of
disorder by introduction of non-magnetic\ Zn$^{2+}$ $(S=0)$ impurities at the
Cu$^{2+}$ site induces long-range AF\ ordering at the Neel temperature
$T_{\text{N}}$ ranging from 3 to 8 K for 0.01 $\leq$ $x$ $\leq$
0.08.\cite{azuma1997} The specific heat measurements above $T_{\text{N}}$ for
$x=0.02$ and $x=0.04$ also confirmed the gapless antiferromagnetic state from
its linear variation with temperature, which should be observed in gapless
systems such as uniform spin chains.\cite{DCUniform chains} Similar AF
ordering was\ also found in Ni doped SrCu$_{2}$O$_{3}$ from NMR
experiments.\cite{S.Ohsugi} On the other hand, in the systems which are beyond
the QCP, i.e., on the LRO side of the phase diagram such as the 2D square
planar La$_{2}$CuO$_{4},$ doping with Zn leads to a\ reduction of
$T_{\text{N}}.$\cite{chakraborthy} In contrast, the physics of defect induced
moments in \ geometrically frustrated spin systems, such as
kagome,\cite{Olariu}$^{,}$\cite{Dommange} hyper-kagome systems\cite{Na4Ir3O8},
etc. is quite different from non-frustrated spin systems. These interesting
results inspired us to look for new quantum spin systems and investigate the
impurity effects in them.

Recently, we have reported the magnetic susceptibility and the specific heat
of a new spin-ladder system BiCu$_{2}$PO$_{6}$, in which the exchange
interaction along the leg $J_{1}/k_{B}$ was found to be $80$ K and the ratio
of the rung to leg exchange coupling $J_{2}/J_{1}$ to be nearly $1$%
.\cite{ours} From these exchange\ couplings and magnetic specific heat, we
estimated a spin-gap ($\Delta/k_{B}$) of $34$ K, which is in good agreement
with previously reported value of $33$ K.\cite{OliverJACS} The
first-principles\ band structure and magnetic susceptibility
analysis\cite{ours} suggested a strong inter-ladder coupling $J_{3}/J_{1}%
\sim0.75$ in the $bc-$plane and a\ next-nearest-neighbor $(nnn)$ exchange
interaction along the leg\ $J_{4}/J_{1}\sim0.34.$ While we are not aware of
any calculations of the QCP for 2D coupled two-leg ladders with a frustrating
$nnn$ interaction along the leg, $J_{3}/J_{1}\sim0.75$ would seem to be very
large and presumably close to a QCP. Perturbing this ground state by doping
with non-magnetic or magnetic impurities can reveal important information
about the nature of the system.

Furthermore in an earlier paper,\cite{msrBobroff} we have used local probe
measurements (nuclear magnetic resonance NMR and muon spin resonance $\mu$SR)
to probe impurity (Zn and Ni) effects in the BiCu$_{2}$PO$_{6}$ spin-ladder.
\ We have shown the appearance of impurity-induced moments leading eventually
to spin-freezing at low temperatures. \ We now report bulk (magnetic
susceptibility and heat capacity) measurements on pure, and doped (homovalent
Zn, Ni at Cu site and heterovalent Pb, Ca at Bi site) HAF spin-ladder
BiCu$_{2}$PO$_{6}$ i.e. Bi(Cu$_{1-x}$Zn$_{x}$)$_{2}$PO$_{6},$ Bi(Cu$_{1-y}%
$Ni$_{y}$)$_{2}$PO$_{6}$ and Bi$_{1-z}$Ca$_{z}$Cu$_{2}$PO$_{6}$. While such
measurements validate the sample quality and surely constitute the first steps
to investigate the gross properties of a system such as the occurence of a
spin-gap, phase transitions, etc., they also enable us to reach conclusions
about details such as the applicability (or lack thereof) of \ theoretical
models (isolated spin-ladder, Majumdar-Ghosh chain, dimer model, etc.) and the
value of the impurity-induced magnetic moment. \ These are useful inputs to
theorists to refine and improve their theoretical models. \ Our magnetic
susceptibility measurements on Zn-doped BiCu$_{2}$PO$_{6}$ evidence
low-temperature upturns due to induced magnetic moments. \ The data also
suggest that the Zn-induced moment is smaller than that expected for isolated
ladders and is likely related to frustrations present in our system. In the
case of Ni doping, a transition to a spin-glass-like disordered state is
clearly evident in the susceptibility data. Anomalies were also observed in
the heat capacity data. Similar to Zn-doped samples, in Ni-doped samples the
induced moment (combination of the Ni magnetic moment and the neighbouring
staggered moments) is found to be smaller than that for an isolated $S=1$
entity. No anomalies were found in the susceptibility or heat capacity data of
Ca or Pb doped samples.

\section{\textbf{EXPERIMENTAL DETAILS}}

Polycrystalline samples of BiCu$_{2(1-x)}$Zn$_{2x}$PO$_{6}$, $(0\leq x\leq1)
$, BiCu$_{2(1-y)}$Ni$_{2y}$PO$_{6}$, $(0\leq y\leq0.2)$, Bi$_{1-z}$Pb$_{z}%
$Cu$_{2}$PO$_{6},(0\leq z\leq0.07)$ and Bi$_{1-z}$Ca$_{z}$Cu$_{2}$%
PO$_{6},(0\leq z\leq0.15)$ were prepared by conventional solid state reaction
methods, using Bi$_{2}$O$_{3}$, CuO, ZnO, NiO, CaCO$_{3}$ and (NH$_{4}$)$_{2}
$HPO$_{4}$. Samples were ground, pelletized, and fired at 770$%
{{}^\circ}%
$C for four days with three intermediate grindings. X-ray diffraction
measurements were performed using a\ PANalytical X'pert PRO powder
diffractometer. All the above samples were found to be\ single-phase and
structural parameters were refined using Fullprof software.\cite{fullprof}
Magnetization $M$ was measured as a function applied magnetic field $H$ and
temperature $T$ $\ (2\leq T\leq300$ K$)$ using a SQUID magnetometer and a
vibrating sample magnetometer (VSM) of a Physical Property Measurement System
(PPMS), both from Quantum Design, Inc. The specific heat $C_{p}$ measurements
were performed as a function of \ $T$\ in the range 1.8 to 70 K by the
relaxation method\ using the PPMS.

\section{\textbf{Results and Analysis}}

\subsection{Crystal structure and lattice parameters}

In the crystal structure of BiCu$_{2}$PO$_{6}$,\cite{FAbhraham} two
edge-shared CuO$_{5}$ distorted square pyramids, which are built from two
different Cu (Cu1 and Cu2) and O (O1 and O2) atoms, respectively, form zigzag
two-leg ladders along the crystallographic $b-$direction. The bond angle and
bond length along the leg (Cu1-O1-Cu2) of the ladder are about 112$^{\circ}$
and 3.2 \AA ,\ respectively and along the rung (Cu1-O2-Cu2) the corresponding
values are 92$^{\circ}$ and 2.8 \AA ,\ respectively. The Bi atoms are
positioned between the ladders in the $bc-$plane as shown in Fig. 1(a). These
planes are separated by PO$_{4}$ tetrahedron units. Various exchange couplings
$(J_{1},J_{2},J_{3},$ and $J_{4})$ between Cu atoms in its $bc$-plane have
been indicated in Fig. 1(b).

All the above mentioned samples were found to be single phase from x-ray
diffraction (xrd) measurements and Rietveldt refinement analysis. The
Rietveldt refinement pattern for the undoped sample is shown in Fig. 2(a). The
xrd patterns for Zn doped samples were also indexed based on the orthorhombic
space group $Pnma$. The xrd\ peaks shift towards smaller angles $2\theta$ as
the Zn content changes from $x=0.0$ to $x=1.0$ which is a clear indication of
a change in the lattice parameters. The yielded lattice parameters $a$, $b,$
and $c$ from the refinement are plotted as a function of $x$ in Fig. $2(c)$.
The lattice parameters of BiCu$_{2}$PO$_{6}$ ($x=0$) and BiZn$_{2}$PO$_{6}$
($x=1$) are in reasonable agreement with previously reported
data\cite{FAbhraham} and the values for intermediate concentrations also
smoothly interpolate between the end points. The changes in lattice parameters
on Ni, Pb, and Ca doping (not shown here) are somewhat less than that on Zn
doping .

\section{\bigskip Magnetic susceptibility and heat capacity}

\subsection{\bigskip\textbf{Undoped BiCu}$_{\mathbf{2}}$\textbf{PO}%
$_{\mathbf{6}}$\textbf{\ }}

The temperature$\ $dependence of bulk magnetic susceptibility $\chi(T)=M/H$
for BiCu$_{2}$PO$_{6}$ is shown in Fig. 3. The $\chi(T)$ shows a broad maximum
at about $56$ K, indicative of a\ low-dimensional magnetic system, thereafter
decreasing steeply with decreasing temperature. This is indicative of a gap in
the spin excitation spectrum in agreement with the 2-leg ladder geometry
suggested by the structure. A small upturn appears below $7$ K, likely due to
extrinsic paramagnetic impurities and /or natural chain breaks in our
polycrystalline samples. We first fit the low-temperature part of $\chi(T)$
below 15 K to the\ equation\cite{troyer 1994}%

\begin{equation}
\chi(T)=\chi_{_{0}}+\frac{C}{(T-\theta)}+\frac{A}{\sqrt{T}}\exp\left(
-\frac{\Delta}{k_{\text{B}}T}\right) \label{gap eqn}%
\end{equation}

This yields the parameters $\chi_{_{0}}=(4.4\pm0.2)\times10^{-4}$ cm$^{3}$
/mole Cu$,$ $\ C=(3.6\pm0.2)\times10^{-4}$ cm$^{3}$ K /mole Cu$,$ $\theta
\sim0$ K$,$ $A=\frac{Ng^{2}\mu_{B}^{2}}{2k_{B}\sqrt{\pi\left(  \frac{\gamma
}{k_{B}}\right)  }}=(2.6\pm0.1)\times10^{-2}$ and the spin-gap $\Delta
/k_{B}=(45\pm3)$ K$.$ Here $k_{B}$ is the Boltzmann constant, $N$ is the
Avogadro number, $g$ is the Lande$%
\acute{}%
$ $g$-factor (fixed to $2.1$, a typical value in cuprates), and $\gamma$ is
the curvature of\ magnon dispersion at the band minimum. The temperature
independent susceptibility $\chi_{_{0}}$ is the sum of the core diamagnetic
susceptibility $\chi_{core}$, the Van Vleck susceptibility $\chi_{vv}$ and a
residual susceptibility $\chi_{res}$ (this will be explained shortly). The
$\chi_{core}$ is calculated\cite{P.W. Selwood} to be $-0.6\times10^{-4} $
cm$^{3}$/mol Cu. Our recent NMR results\cite{Libu Alexander} have shown that
the $^{31}$P NMR shift (and hence the spin susceptibility) falls exponentially
at low-$T$ due to the gap, there remains a non-zero NMR shift as
$T\rightarrow0$. \ This amounts to a residual spin susceptibility $\chi
_{res}=1.4\times10^{-4}$ cm$^{3}$/mol Cu. Using this value of $\chi_{res}$\ we
get $\chi_{vv}=3.6\times10^{-4}$ cm$^{3}$/mol Cu which is somewhat higher than
$\chi_{vv}$ of other cuprates. The Curie constant corresponds to about $0.1$
\% of isolated spin-%
${\frac12}$
impurities. In a latter section, we analyse the impurity-induced
susceptibility in greater detail. The ratio $\gamma/\Delta$ was also found
from the fit. \ However, due to the small amount of data at low-$T$, the
result was found to be dependent on the $T$-range used for the fit and is
therefore considered unreliable. \bigskip

In order to gain further insight into the magnetic interactions present in
this system, we now attempt to fit the susceptibility data in the full
temperature range for undoped BiCu$_{2}$PO$_{6}$. \ Note that while the
measured susceptibility has a low-$T$ Curie term (albeit small) the spin
susceptibility obtained via the $^{31}$P\ NMR shift measurements
($\chi_{\text{NMR}}(T)$, detailed in a separate paper\cite{Libu Alexander}) is
intrinsic and has no Curie term. \ We have, therefore, used $\chi_{\text{NMR}%
}\ (T)$ to carry out the analysis following Equation (\ref{coupled ladder})
below.\cite{D.C Johnston} This kind of\ equation has been used in the past to
determine the strength of inter-dimer, inter-chain, and inter-ladder
interactions in the corresponding systems (non-frustrated) like SrCu$_{2}%
$O$_{3}$,\cite{DC1996} CaV$_{2}$O$_{5}$,\cite{D.C Johnston}, Sr$_{3}$Cr$_{2}%
$O$_{8}$,\cite{Sr3Cr2O8} etc.%

\begin{equation}
\chi_{\text{NMR}}\ (T)=\frac{Ng^{2}\mu_{B}^{2}}{J_{1}}\left(  \frac
{\chi_{\text{ladder}}^{\ast}\left(  J_{1,}J_{2},T\right)  }{1+\lambda
\chi_{\text{ladder}}^{\ast}\left(  J_{1,}J_{2},T\right)  }\right)
\label{coupled ladder}%
\end{equation}

Here, $\chi_{\text{ladder}}^{\ast}\left(  J_{1,}J_{2},T\right)  =\frac{J_{1}%
}{Ng^{2}\mu_{B}^{2}}\chi_{\text{ladder}}(T)$ (with $\chi_{\text{ladder}}(T)$
being the isolated ladder susceptibility\cite{D.C Johnston}) while $\lambda$
$(=2(z_{eff}-z_{0}))$ accounts for the intersystem exchange coupling. The
fitting parameters depend somewhat on the $T$-range used for the fit. \ The
obtained values for the leg and rung coupling are about $J_{1}/k_{B}\sim80$ K
$\&$ $J_{2}/k_{B}=78$ K. \ Also, $\lambda$ is found to be in the range $8-10$.
Given that $z_{0},$the magnetic coordination number, in the case of the
isolated ladder is $3$, we get the effective magnetic coordination number
$z_{eff}$ $\sim7-8$. \ \ Using this and our experimental value of $\chi
^{max}=\chi_{\text{NMR}}^{max}$ $=1.25\times10^{-3}$ cm$^{3}$/mole Cu, we
get$\frac{\chi^{max}J^{max}z_{eff}}{Ng^{2}\mu_{B}^{2}}=0.42-0.48$ which lies a
little above the universal curve (for nonfrustrated bipartite AF spin
lattices) given by Johnston (see Fig. 4 of Ref. \cite{DC1996}). \ Further,
\ Johnston\cite{DC1996} had made the observation that $\frac{\chi^{max}%
J^{max}z_{eff}}{Ng^{2}\mu_{B}^{2}}$ lies above the universal curve for highly
frustrated systems like the triangular system and other cluster systems where
the frustrating interaction is equal to the other interaction. \ One might,
therefore, naively conclude that BiCu$_{2}$PO$_{6}$ might be only moderately frustrated.

The plot of$\ \ \chi_{spin}T$ $(=(\chi-\chi_{core}-\chi_{vv})T)$ vs $T$ for
undoped BiCu$_{2}$PO$_{6}$ is shown in Fig. 3(b). Even at $300$ K, which is
nearly four times the value of $J^{\max}/k_{B}$, the value of $\chi_{spin}T$
is about $0.185$ and smaller than $0.375$, which is the value of $\chi
_{spin}T$ in the paramagnetic limit. This suggests that the system is not in
the paramagnetic limit even at $300$ K. As a further illustration, we show the
spin magnetization $M_{spin}(H)$ as a function of applied field at $T=350$ K
with the comparison of Brillouin function for $S=%
\frac12
$ and $g=2.1$ in the inset of Fig. 3(b). \ Clearly, the data are well below
the Brillouin function curve. These results suggest that AF and frustrating
interactions are present even at $T=350$ K.

\bigskip

Theoretically, a coupled two-leg spin ladder system ($J_{1}=J_{2}$) has a
quantum critical point (QCP) at about $J_{3}/J_{1}=0.3$, which separates the
spin-gap region from the LRO antiferromagnetic region.\cite{sachdevsci2000}
\ Experimentally, however, we have proven that this system has a spin gap of
about $35-40$ K. Recent inelastic neutron scattering results on BiCu$_{2}%
$PO$_{6}$ yield a spin-gap of about $1.5$ meV (equivalent to $20$ K).
\cite{adroja} The difference with our results could be due to a wave-vector
dependence of the spin-gap affecting the neutron scattering results on powder
samples. \ From theory it is known that in two-leg ladders the existing
spin-gap increases due to frustrating $nnn$ interactions.\cite{SR White}Also,
from theoretical DMRG calculations, it is predicted that the three-leg spin
ladder (which is gapless) will develop a spin-gap in case of frustrating $nnn$
interactions. The continued presence of the spin-gap state in our system
inspite of the relatively large inter-ladder couplings (predicted from
electronic structure calculations\cite{ours}) points to the presence of
frustrating interactions. It is likely that the next-nearest-neighbour ($nnn$)
coupling $J_{4}$ is frustrating and plays a role in sustaining the spin-gap
state. \ 

Specific heat $C_{p}(T)$ of BiCu$_{2}$PO$_{6}$ is shown in Fig. 4. \ We have
analyzed $C_{p}(T)$ of BiZn$_{2}$PO$_{6}$, the non-magnetic analog of
BiCu$_{2}$PO$_{6}$, using the Debye model.\cite{ESR Gopal} The data couldn't
be fit using a single Debye temperature, probably because of different atoms
having different atomic weights. We were able to fit the data to the following
formula which contains a linear combination of two Debye integrals.%

\begin{equation}
C_{p}(T)=9rNk_{B}%
{\displaystyle\sum\limits_{i=1,2}}
C_{i}\left(  \frac{T}{\theta_{D}^{i}}\right)  ^{3}%
{\displaystyle\int\limits_{0}^{x_{D}^{i}}}
\frac{x^{4}e^{x}}{(e^{x}-1)^{2}}dx\label{debye}%
\end{equation}

Here $r$ is the number of atoms per formula unit, $\theta_{D}^{i}$ is a Debye
temperature and $x_{D}^{i}=\theta_{D}^{i}/T$ . \ The\ obtained parameters are
$C_{1}=0.20\pm0.01,$ $\theta_{D}^{1}=183\pm2$ K$,$ $C_{2}=0.41\pm0.01,$ and
$\theta_{D}^{2}=456\pm6$ K. Further we also fit the $C_{p}(T)$ of BiZn$_{2}%
$PO$_{6}$ data to the low$-T$ Debye expression ($\frac{12\pi^{4}}{5}%
Nrk_{B}\left(  \frac{T}{\theta_{D}}\right)  ^{3}$) in the $T-$range from $2$
to $20$ K. The yielded $\theta_{D}$ is ($325\pm2)$ K and is intermediate to
$\theta_{D}^{1}$ and $\theta_{D}^{2}.$ In the same $T-$range, we go ahead and
fit the $C_{p}(T)$\ data of BiCu$_{2}$PO$_{6}$ using the following equation%

\begin{equation}
C_{p}(T)=\frac{12\pi^{4}}{5}Nrk_{B}\left(  \frac{T}{\theta_{D}}\right)
^{3}+\frac{3}{2}Nrk_{B}\left(  \frac{\Delta}{\pi\gamma}\right)  ^{1/2}\left(
\frac{\Delta}{k_{B}T}\right)  ^{3/2}\left[  1+\frac{k_{B}T}{\Delta
}+0.75\left(  \frac{k_{B}T}{\Delta}\right)  ^{2}\right]  \exp\left(
\frac{-\Delta}{k_{B}T}\right) \label{gap heat capacity}%
\end{equation}

Here the yielded value of $\theta_{D}=330\pm2$ from the first part is nearly
the same as the value of $\theta_{D}$ of BiZn$_{2}$PO$_{6}$. Hence we used
$C_{p}(T)$ of BiZn$_{2}$PO$_{6}$ as the lattice contribution for the
BiCu$_{2}$PO$_{6}$ (as mentioned before) and the magnetic part\ $\frac
{C_{m}(T)}{T}$ $(x=0.0)$ is plotted in Fig. 6. Exponential behavior, a broad
maximum, and a lack of LRO are the indications of spin-gap systems. The second
term is the magnetic specific heat for two-leg spin ladders.\cite{D.C
Johnston} A fit of our data to the above Equation yields $\Delta=43\pm2$ K (in
good agreement with the $45$ K value from low-$T$ susceptibility analysis) and
$\gamma/\Delta=6\pm1$. The magnetic entropy $S_{m}$ is estimated by
integrating the data $\frac{C_{m}(T)}{T}$ vs $T$. The normalized $S_{m}$
$(S_{m}^{\ast})$ is plotted as a function of reduced temperature $T^{\ast
}(=k_{B}T/J_{1})$ in the inset of Fig. 6. The value of $S_{m}^{\ast}$ at
$T^{\ast}=0.5$\ is $0.2$ which is $35$\% of the value of $0.56$ for an $S=%
\frac12
$ uniform chain which is non-frustrated.\cite{johnstonchains} The quenching of
magnetic entropy at low temperature is a hallmark of frustrated spin systems.

\subsection{\bigskip\textbf{Zn-doped BiCu}$_{\mathbf{2}}$\textbf{PO}%
$_{\mathbf{6}}$}

The magnetic susceptibilities of Zn$^{2+}$ $(S=0)$ doped at Cu$^{2+}$ site in
the\ BiCu$_{2}$PO$_{6}$ samples i.e., BiCu$_{2(1-x)}$Zn$_{2x}$PO$_{6},$
$(0\leq x\leq0.5)$, were measured in an applied field of 5000 Oe as shown in
Fig. 5(a). The susceptibility \ at low temperatures increases with the doping
concentration $x$. Introduction of Zn$^{2+}$ at Cu$^{2+}$ in the
ladder\ creates spin-vacancies, which are expected to induce localized
magnetic moments on the neighboring Cu sites. \ The effect of induced moments
has certainly been observed by us in NMR and $\mu$SR
measurements.\cite{msrBobroff} For the case of an isolated $S=1/2$ HAF 2-leg
ladder, the effective magnetic moment per vacancy is expected to be exactly
that due to an $S=1/2$ entity. For $x\geq0.02$, it is clear that the low$-T$
$(T<6$ K$)$ behavior is not Curie-like (there is a deviation from linearity in
the log-log plot of Fig. 5(a)). We analyse the low-$T$ behaviour in the
susceptibility in the following manner. There are both intrinsic (natural
defects/chain breaks, impurity-induced effects, spin-gap) and extrinsic
(impurity phase) contributions to the susceptibility in all the samples. \ As
argued,\cite{Libu Alexander} the spin susceptibility due only to the spin-gap
($\chi_{\text{NMR}}$) is obtained from the $^{31}$P\ NMR shift. \ Next, we
subtract $\chi_{\text{NMR}}$ from the measured susceptibility $\chi$. This
yields $\chi_{\text{imp}}$($x,T$) which is the combination of a $T$%
-independent part and a $T$-dependent part due to defects, dopants, and
extrinsic impurities and is shown in Fig. 5(b). \ As seen, the undoped sample
has a weakly $T$-dependent contribution. We then plot $\chi_{\text{imp}%
}^{\text{corr}}(T$ $+\theta)$ divided by the Zn concentration, as a function
of $T$ in the inset of Fig. 5(b). \ The impurity susceptibility is
$\chi_{\text{imp}}^{\text{corr}}=$ $\chi_{\text{imp}}(x)-\chi_{\text{imp}%
}(x=0)-\chi_{0}$ where $\chi_{0}$ is a $T$-independent term. \ Here, $\theta
$\ is taken equal to $T_{g}$ which is the temperature below which a frozen
magnetic state was evidenced from $\mu$SR measurements. For a pure Curie-Weiss
behaviour, $\chi_{\text{imp}}^{\text{corr}}(T+\theta)$ should be
$T$-independent giving the value of the Curie constant. Our result is seen to
be not quite $T$-independent (the nature of the data are not much dependent
for $\theta$ between $0$ and $T_{g}$). \ However, $\chi_{\text{imp}}$
increases linearly with Zn content. \ \ Also, it is seen that the
impurity-induced susceptibility is smaller (consistent with an effective
paramagnetic moment of $S\sim0.3$) than the case where each Zn induces an
effective $S=1/2$ moment (the dashed line in the inset of Fig. 5(b)). \ This
might be related to the presence of frustrating interactions in the compound.
While the details of $^{31}$P NMR measurements in pure and doped BiCu$_{2}%
$PO$_{6}$ are discussed in a separate paper,\cite{Libu Alexander} suffice to
say at this point that the data are consistent with the nominal Zn/Ni content
in the ladder. \ The position of the broad maximum in $\chi(T) $ does not
appear to change with Zn content as shown in Fig. 5(a). Surprisingly even for
$x=0.5$, a broad maximum is seen. No difference in the zero-field-cooled (ZFC)
and field-cooled (FC) susceptibilities was seen at low$-T$. However, the
deviation from the Curie-behavior seems to suggest transition to a new
magnetic state. These results are different from those of SrCu$_{2}$O$_{3}$,
in which cusp like anomalies were found in $\chi(T).$\cite{azuma1997}

\bigskip

Quite generally, the replacement of Cu$^{2+}$ ion with Zn$^{2+}$ in a
spin-liquid with a singlet ground state like in HAF\ two-leg spin ladders
induces an effective spin-%
${\frac12}$
moment.\cite{awsandwik} However, a deviation from the free, \ spin-%
${\frac12}$
moment was predicted in $S=0$ doped, one-dimensional, dimerized, frustrated
(Majumdar-Ghosh) chains.\cite{Normand} Considering that BiCu$_{2}$PO$_{6}$ can
be thought of as consisting, essentially, of coupled Majumdar-Ghosh chains,
predictions by Normand and Mila might be valid here. \ The reduced Curie
constants (compared to the $S=1/2$ value) observed by us in Zn-doped samples
might be an evidence of this. \ However, there may be other explanations.
\ For higher doping concentrations, interaction between the induced moments
result in deviations in the $\chi(T)$ from a Curie law. The origin of the
disordered magnetic state which has been observed may be connected to random
exchange interactions between impurity induced local magnetic moments on Cu
sites on ladders in different $bc$-planes.

In order to better understand the properties of Zn doped BiCu$_{2}$PO$_{6}$,
we have also performed the specific heat $C_{p}(T)$ measurements for
$x\leq0.30.$ After subtracting the lattice contribution from the data for the
non-magnetic analog BiZn$_{2}$PO$_{6}$, the obtained magnetic contribution
$C_{m}/T$ versus $T$ is plotted in Fig. 6. Here we found that even for small
doping level $x=0.01$, an upturn is seen at low-$T$. This is possibly
indicative of an onset of magnetic freezing. For higher $x$, broad anomaly is
seen around 4 K. \ These suggest the loss of the spin-gap and a freezing of
impurity-induced moments. \ Due to the small magnetic heat capacity and the
weak anomaly, we don't have good accuracy in determining the freezing
temperature. \ However, the anomalies in $C_{m}$/$T$ vs $T$ are in the same
range as found by our $\mu$SR measurements on the same
samples.\cite{msrBobroff}

\subsection{\textbf{Ni-doped BiCu}$_{\mathbf{2}}$\textbf{PO}$_{\mathbf{6}}$}

The $\chi(T)$ of Ni$^{2+}$ $(S=1)$\ doped BiCu$_{2}$PO$_{6}$ were measured in
the $T-$range $2$ to $300$ K and are shown in Fig. 7(a). At low$-T,$ an
enhancement of susceptibility with Ni content $(y)$ was observed and for
higher contents $(y\geq0.15)$, even the high$-T$ susceptibility increased with
$y$. At low doping levels, we again carried out the analysis as for Zn doping.
\ The plot of $\chi_{\text{imp}}^{\text{corr}}(T+\theta)$ for Ni-doped samples
is shown in Fig. 7(b). \ A Ni dopant is expected to give rise to two
contributions; (i) a staggered magnetisation on the neighbouring Cu sites as
for the Zn and (ii) the magnetic moment of the Ni itself. \ While the
"effective" Curie constant induced per percent of Ni (about $0.006$ cm$^{3}%
$K/mole) is clearly greater than that for Zn, the value is smaller than that
expected for $1$\% of paramagnetic $S=1$ entities. This indicates that the
Ni$^{2+}$ impurity moments may be coupled antiferromagnetically\ with induced
moments on neighboring Cu sites. However, it is also possible that there is a
ferromagnetic coupling between the Ni and the neighbouring Cu. \ An agreement
with our experimental results can be achieved if there is as well a
frustration related screening which decreases the size of the paramagnetic
cloud comprising of the Ni moment and the induced moments on the neighbouring
Cu. \ Similar to Zn doped samples, at low temperatures $\chi(T)$ deviates
significantly from a Curie behavior. In addition, there are\ differences
observed in ZFC and FC $\chi(T)$ data measured in an applied field 100 Oe as
shown in inset of Fig. 7. Specific heat data are further measured to
understand these anomalies. Sharp\ cusp like anomalies are found with Ni
doping as shown in Fig. 8. The temperatures at which these anomalies are
present are similar to those in $\chi(T)$ data. The exponential $T-$dependence
and the broad maximum present in the $C_{m}(T)$ of undoped composition were
lost on Ni-doping indicating a loss of the spin-gap. \ The cusps in the heat
capacity are likely associated with spin-freezing (similar features have been
noticed in Zn- and Ni-doped SrCu$_{2}$O$_{3}$ [Ref. \cite{azuma1997}] and
CuGeO$_{3}$ [Ref. \cite{Grenier98}]). \ Note that the susceptibility or heat
capacity anomalies are not sharply defined in our data (in fact, if one looks
at $C_{m}$ vs $T$, only a broad feature is seen). Keeping this in mind, in our
$\mu$SR measurements (where the freezing temperature is obtained precisely) on
the same samples,\cite{msrBobroff} we observed spin-freezing at similar
temperatures as the heat capacity anomalies.

\subsection{Ca/Pb-doped \textbf{BiCu}$_{\mathbf{2}}$\textbf{PO}$_{\mathbf{6}}
$}

The replacement of Ca$^{2+}/$Pb$^{2+}$ in place of Bi$^{3+}$ should generate
holes. The $\chi$ of hole-doped samples Bi$_{1-z}$(Pb/Ca)$_{z}$Cu$_{2}$%
PO$_{6}$ $(0\leq z\leq0.15)$ are shown in Fig. 9(a). \ The $\chi$ increases
with doping concentration $(z)$ and no anomalies were found at low-$T$, in
contrast with Zn/Ni doped samples. Specific heat data are also consistent with
this, as shown in the inset of\ Fig. 9. There is no change in the position
of\ broad maximum with doping. To obtain Curie constant of Ca doped samples,
we again plotted $\chi_{\text{imp}}^{\text{corr}}T$ as a function of $T$ in
Fig. 9b. For clarity, we have not shown the data of Pb doped samples in Fig.
9(b). The effective Curie constant per percent of doped Ca/Pb amounts to less
than one-fourth of that due to an $S=$ $%
\frac12
$ moment for each hole, which is even smaller than the value obtained for Zn
doped samples. \ This might be expected since Pb or Ca substitution at the Bi
site does not alter the ladder geometry but is rather expected to dope holes.
\ It is not known what type of spin-state these holes (which are evidently
localised) form due to hybridisation of the Pb/Ca orbitals with the
neighbouring orbitals. The Curie term might then be partly due to oxygen
non-stoichiometry and partly due to the localised holes, apart from any
extrinsic effects.

\ \ \ \ \ \ \ \ \ \ \ \ \ \ \ \ \ \ \ 

\section{CONCLUSION}

In summary, we have given a detailed account of preparation and properties of
Ni-, Zn-, Pb-, and Ca-doped BiCu$_{2}$PO$_{6}$. \ While Zn has complete solid
solubility, Ni can be substituted upto a concentration of about $20$ \% and Ca
up to about $15$\%. \ \ The effective magnetic coordination number for
BiCu$_{2}$PO$_{6}$ was found to be about $7$ from an analysis of the
susceptibility data. \ The heat capacity for BiZn$_{2}$PO$_{6}$ is well fit in
the $T$-range $2$ K-$100$ K by a superposition of two Debye functions with
Debye temperatures of about $180$ K and $450$ K. \ The magnetic entropy of
BiCu$_{2}$PO$_{6}$ is quenched at low temperatures, presumably due to
frustration. \ Both, Ni- and Zn-doping in BiCu$_{2}$PO$_{6}$ are found to give
rise to static, disordered magnetism at low temperatures. \ While clear
anomalies were found in the temperature dependence of magnetic susceptibility
and magnetic heat capacity of the Ni-doped samples, the Zn-doped samples
exhibited merely a deviation from Curie behaviour at low temperature while the
heat capacity showed a weak anomaly. \ \ The impurity-induced moment is found
to be reduced from the $S=1/2$ value in Zn-doped BiCu$_{2}$PO$_{6}$ which
might be due to frustration effects.\cite{Normand} On the other hand, Ca- and
Pb-doped samples gave rise to only weak Curie terms and a nearly unchanged
magnetic heat capacity $C_{m}(T)$ which implies that the magnetic
ladders/planes are not disturbed by the doping at Bi site and any carriers
generated localise at low temperatures.

\begin{acknowledgement}
We thank the Indo-French Center for the Promotion of Advanced Research and
ARCUS Ile de France-Inde.
\end{acknowledgement}

\textbf{Figure Captions}

FIG. 1 (Color Online) (a)The crystallographic\ $bc-$plane of BiCu$_{2}$%
PO$_{6}$ is shown and it can be seen that coupled two-leg zigzag ladders
formed by Cu (blue) \& O (red) atoms and Bi (green) atoms are positioned in
between. The various major\ exchange interactions between Cu atoms are shown
in the schematic Fig. 1(b).

FIG. 2 (Color Online) The x-ray diffraction pattern of BiCu$_{2}$PO$_{6}$ is
shown along with the Rietveldt refinement. The black open circles indicate the
experimental xrd\ data, the solid red\ line is calculated based on the
refinement, green vertical short lines are the Bragg positions and blue
line\ is the difference of measured and calculated data. The refinement
parameters are R$_{p}$= $14.0$, R$_{wp}$ = $11.0$ and $\chi^{2}=4.2.$ (b) The
xrd patterns of BiCu$_{2(1-x)}$Zn$_{x}$PO$_{6}$ $(x=0.0,$ $0.5$ $\&$ $1.0)$
with hkl-indices are shown. (c) The lattice constants $a$, $b$, and $c$ as
a\ function of Zn\ doping content ($x)$ are shown in Fig. 2c.

FIG. 3 (Color Online) (a)Magnetic susceptibility $\chi(T)=$($M/H$) vs.
temperature $T$ \ (open squares) for BiCu$_{2}$PO$_{6}$ in an applied field of
$5000$ Oe. Also plotted is $\chi_{\text{NMR}}(T)$ (open triangles). \ The
solid line is a fit to Equation (\ref{coupled ladder}) of $\chi_{\text{NMR}%
}(T)$. \ As shown in the\ inset Fig. 3(a),\ dashed line is a fit to the
Equation (\ref{gap eqn}) as explained in the text. In Fig. 3(b), $\chi
_{spin}T$ is plotted as a function of $T$ and inset shows magnetization $(M)$
as a function of magnetic field $(H)$ upto 90 kOe and a comparison with the
Brillouin function at 350 K. The arrow mark indicates the value corresponding
to the Curie constant ($C$) for free spin-$%
\frac12
$.

FIG. 4 (Color Online) (a) The specific heat $C_{p}$ as a function of $T$ for
BiCu$_{2}$PO$_{6}$ (blue open up-triangles) and BiZn$_{2}$PO$_{6}$\ (black
open\ down-triangles) is shown and fitted with the equations mentioned in the
text. Inset shows the data of$\ C_{p}$ as a function of $T^{3}$ for BiZn$_{2}%
$PO$_{6}$. (b) The magnetic contribution to the specific heat of BiCu$_{2}%
$PO$_{6}$ (left axis) is plotted vs. the normalised temperature $k_{\text{B}%
}T/J_{1}$. \ Also shown is the magnetic entropy $S_{\text{M}}$ (right axis)
vs. $k_{\text{B}}T/J_{1}$.

FIG. 5(a) (Color Online) Magnetic susceptibility vs. temperature $\chi(T)$ for
BiCu$_{2(1-x)}$Zn$_{2x}$PO$_{6}$, ($0\leq x<0.5$) in the range 2 K to 300 K
measured in the applied field 5000 Oe. (b) A plot of the non-spin-gap part of
the susceptibility $\chi_{\text{imp}}$for BiCu$_{2(1-x)}$Zn$_{2x}$PO$_{6}$
(i.e., only from impurity-induced effects and from extrinsic impurities) is
shown as a function of $T$. \ The inset shows $\chi_{\text{imp}}^{\text{corr}%
}(T+\theta)$ per percent Zn as a function of $T$. The dashed horizontal line
represents the value corresponding to an $S=1/2$ induced moment per percent
Zn. \ 

FIG. 6 (Color Online) Magnetic specific heat divided by temperature
$C_{m}(T)/T$ vs. $T$ for BiCu$_{2(1-x)}$Zn$_{2x}$PO$_{6}$, ($0\leq x\leq0.3$)
in the range 2 K to 50 K.

FIG. 7(a) (Color Online) The $\chi(T)$ of BiCu$_{2(1-y)}$Ni$_{2y}$PO$_{6}$
$(0\leq y\leq0.2)$, measured in a magnetic field of 5000 Oe and plotted on a
log-log scale. The\ ZFC and FC susceptibility $\chi(T)$ measured in 100 Oe is
shown in the inset. The arrow marks at bifurcation points are indicative of
the freezing temperature. (b) A plot of the non-spin-gap part of the
susceptibility $\chi_{\text{imp}}$for BiCu$_{2(1-x)}$Ni$_{2x}$PO$_{6}$ (i.e.,
only from impurity-induced effects and from extrinsic impurities) is shown as
a function of $T$. \ The inset shows \ $\chi_{\text{imp}}^{\text{corr}%
}(T+\theta)$ per percent Ni as a function of $T$. The dashed horizontal line
represents the value corresponding to an $S=1$ induced moment per percent Ni.

FIG. 8 (Color Online) Magnetic specific heat divided by temperature
$C_{m}(T)/T$ vs. $T$ for BiCu$_{2(1-y)}$Ni$_{2y}$PO$_{6},(0\leq y\leq0.2),$ in
the range 1.8 to 50 K (plotted on a semi-log scale). Anomalies are seen (arrow
marks) which might be associated with magnetic freezing.

FIG. 9 (Color Online) (a) The $\chi(T)$ of Bi$_{1-z}$(Ca/Pb)$_{z}$Cu$_{2}%
$PO$_{6}$ ($0\leq z\leq0.15$) is shown. Inset shows $C_{m}(T)$ for pure and
Pb-doped ($z=0.07$) samples. (b) A plot of the non-spin-gap part of the
susceptibility $\chi_{\text{imp}}$for Bi$_{1-z}$(Ca/Pb)$_{z}$Cu$_{2}$PO$_{6} $
(i.e., only from impurity-induced effects and from extrinsic impurities) is
shown as a function of $T$. \ The inset shows \ $\chi_{\text{imp}%
}^{\text{corr}}T$ per percent Ca as a function of $T$.

\end{document}